\documentstyle[multicol,epsfig,prl,aps]{revtex}
 
\ifpreprintsty
\def\multb{ }
\def\multe{ }
 \else
\def\multb{ \begin{multicols}{2}}
\def\multe{ \end{multicols}}
 \fi

\begin{document}
\title{Superconductivity of metallic boron in MgB$_{2}$}
\author{J. Kortus$^\dag$, I.I. Mazin$^\dag$, K.D. Belashchenko,$^\ddag$
 V.P. Antropov$^\ddag$, L.L. Boyer$^\dag$}
\address{$^\dag$Center for Computational Materials Science,
Code 6390, Naval Research Laboratory, Washington, DC 20375\\
 $^\ddag$Ames Laboratory, ISU, Ames, IA, 50011 }
\date{\today}
\maketitle

\begin{abstract}
Boron in MgB$_{2}$ forms layers of honeycomb lattices with magnesium 
as a space filler.
Band structure calculations indicate that Mg is substantially
ionized, and the bands at the Fermi level derive mainly from
B orbitals. Strong bonding with an ionic component and
considerable metallic density of states yield a sizeable electron-phonon
coupling. Using the rigid atomic sphere approximation and an analogy to
Al, we estimate the coupling constant $\lambda$ to be of order 1.
Together with high phonon frequencies, which we estimate via
zone-center frozen phonon calculations to be between 300 and 700 cm$^{-1},$
this produces a high critical temperature, consistent with recent
experiments. Thus MgB$_{2}$ can be viewed as 
an analog of the long sought, but still hypothetical, 
superconducting metallic hydrogen.
\end{abstract}
%%\pacs{74.20.Fg, 74.25.J, 71.18.+y, 74.70.Ad}

\multb
Before the discovery of high-temperature superconductors much 
effort was devoted to boosting the critical temperature of conventional,
BCS-Eliashberg superconductors\cite{problema}. 
An exotic and appealing idea going back to the early 60's, 
was that of metallic hydrogen\cite{H}.
The arguments were very simple: due to the light mass, the phonon
frequencies in metallic hydrogen would be very high, of the order
of several thousand K, and the prefactor in the BCS formula would
be very large, so that even a moderate coupling constant 
would provide a sizeable $T_{c}.$ This idea
can be quantified as follows: for monoatomic solids, the
electron-phonon coupling (EPC) constant, $\lambda ,$ which enters the BCS
equation, can be written in the so-called
McMillan-Hopfield form \cite{mm1}, 
$\lambda =N(0)\left\langle I^{2}\right\rangle /M\left\langle 
\omega ^{2}\right\rangle$, 
where $N(0)$ is the density of states (DOS) at the Fermi level 
per spin per atom, $\left\langle I^{2}\right\rangle$ 
is the properly averaged electron-ion matrix element squared, $M$ is the
atomic mass and $\left\langle \omega ^{2}\right\rangle $ is 
(again, properly averaged) the  phonon frequency. 
The product $M\left\langle \omega ^{2}\right\rangle $ 
does not depend on the mass, but on the force constants only\cite{problema}, 
while $\eta =N(0)\left\langle I^{2}\right\rangle ,$
also known as the Hopfield factor, is a purely electronic property.
Correspondingly, light elements, everything else being the same, are
beneficial for superconductivity.

Lacking metallic hydrogen, attention was focused upon compounds with light
elements: carbides, nitrides (arguably, 
the superconductivity in fullerenes was a discovery along this road). 
Indeed, many of them were `high-$T_{c}$
superconductors' on the contemporary scale: 10-15 K. 
It was pointed out\cite{DAP} that 
$\left\langle I^{2}\right\rangle $ is rather large in
these materials due to the relatively high ionicity
(although not as high as in MgB$_{2}$),
but $N(0)$ is rather small. 
This led to the suggestion of cubic MoN,
which would have a larger $N(0)$ than existing
nitrides and carbides, 
as a hypothetical superconductor with $T_{c}>30$ K\cite{MoN}.

The recently discovered medium-$T_{c}$ superconductor
MgB$_{2}$ \cite{expB2Mg} with $T_c\agt$39  K
is clearly a continuation of the same idea. 
The main component, B, is even lighter than C and N. 
Furthermore, electronic structure calculations show
that the compound is not only quite
ionic with a reasonable DOS, but also has strong
covalent B-B bonding (the bonding-antibonding splitting due to
in-plane B-B hopping is about 6 eV)
and thus exhibits strong electron-phonon interactions. 
Interestingly, unlike carbides and nitrides, and similar
to metallic hydrogen, electrons at the Fermi level (and below) are 
predominantly 
B-like. Mg $s$-states are pushed up by the B $p_{z}$ orbitals and
fully donate their electrons to the boron-derived conduction
bands. In the following we will
describe the physics of such `metallic' boron in detail, present an
estimate of the EPC constant, and propose some routes for optimizing 
$T_{c}$ in this kind of compound.

MgB$_{2}$ occurs in the so-called AlB$_{2}$ structure. Borons form a
primitive honeycomb lattice, consisting of graphite-type sheets stacked
with no  displacement. 
The borons form hexagonal prisms with the
base diameter of 3.5\AA\  nearly equal to the height. This
creates large, nearly spherical pores for Mg. As in graphite, the
intraplanar B-B bonds are much shorter than the distance between the
planes, and hence the B-B bonding is strongly anisotropic. 
However, the interplane bonds are only twice as long as 
the intraplane ones, as compared to the ratio of 2.4 in graphite, 
allowing for a significant interplane hopping.

\begin{figure}
\centerline{
\epsfig{file=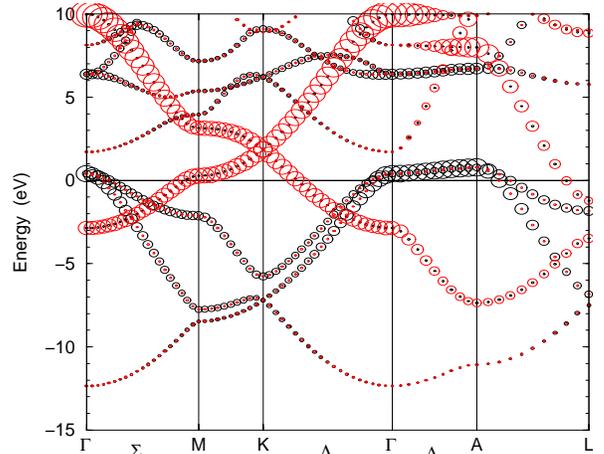,width=0.9\linewidth,clip=true}
}
\setlength{\columnwidth}{\linewidth} \nopagebreak
\caption{Bandstructure of MgB$_2$ with the B p-character.
The radii of the red (black) circles are proportional to the B
 p$_z$ (B p$_{x,y}$) character.}
\label{bands}
\end{figure}

\begin{figure}
\centerline{
\epsfig{file=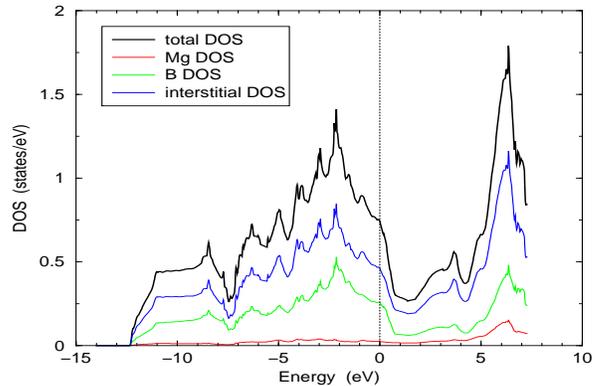,width=0.9\linewidth,height=0.6\linewidth,clip=true}
}
\setlength{\columnwidth}{\linewidth} \nopagebreak
\caption{Total density of states (DOS) and partial DOS for the
MgB$_2$ compound. The small Mg DOS is partially due to the small
$r_{MT}$ of 1.8 $a_B$ used. }
\label{dos}
\end{figure}

We have calculated the electronic structure of MgB$_{2}$ using a general
potential LAPW code\cite{WIEN}. For the Rigid Atomic Spheres calculations we
used the Stuttgart LMTO-TB code\cite{lmto}. For the exchange-correlation
potential, the Generalized Gradient Approximation of Ref.\cite{GGA} was
employed. Despite the rather simple crystal structure, very few
electronic structure calculations for MgB$_{2}$ have been reported
(a model TB calculation of Burdett and Miller\cite{Bur} and a
recent full-potential LMTO study\cite{medv}), and these have concentrated
mainly on chemical bonding, paying hardly any attention to transport
and electronic properties. The results of our LAPW calculations are shown in
Figs.\ \ref{bands} and \ref{dos}. 
We note first that there is almost no valence charge inside the Mg
MT sphere (less than 0.2 $e$).
About half of the total valence charge
resides inside the B spheres, and about the same amount in the
interstitials.
This is partially due to the fact that the chosen LAPW setup
employs rather small MT spheres for Mg.
For the LMTO calculations we used an
atomic sphere of nearly the size of the free Mg atom (up to
3.13$a_{B}),$ and obtained, as expected, a larger charge of 2.8 electrons. 
However less than 25\% of the charge has $s$-character.
The remaining charge of $p$-, $d$-, and $f$- character arises not from Mg
electrons but rather from the tails of the B $p$-orbitals and
contributions from the interstitials.
In fact, one can say that Mg is fully ionized in this compound, however
the electrons donated to the system are not localized on the anion, 
but rather are distributed over the whole crystal. 

\begin{figure}
\centerline{
\epsfig{file=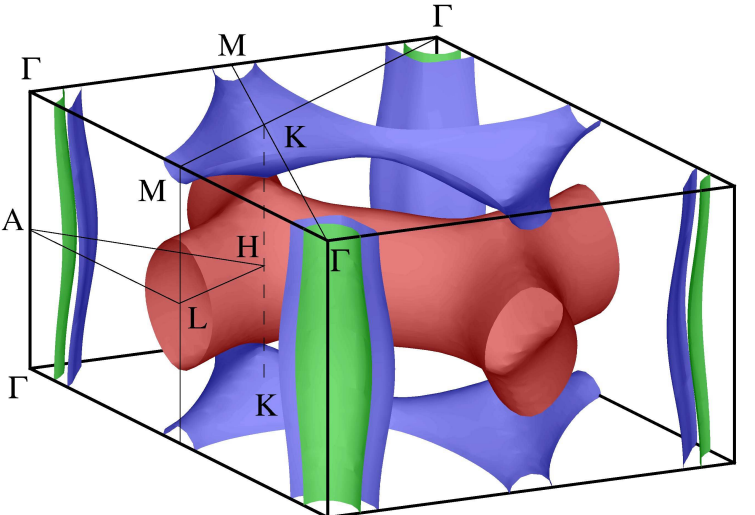,width=0.8\linewidth,angle=-90,clip=true}
}
\setlength{\columnwidth}{\linewidth} \nopagebreak
\caption{The Fermi surface of MgB$_2$. Green and blue cylinders
(hole-like) come from the bonding $p_{x,y}$ bands, the blue tubular network
(hole-like) from the bonding $p_z$ bands, and the red (electron-like)
tubular network from the antibonding $p_z$ band. 
The last two surfaces touch at the K-point.
 }
\label{Fermi}
\end{figure}

The resulting band structure can be easily understood in terms of
the boron sublattice.
We plotted in Fig.\ \ref{bands} the character of the bands (we show only the B
$p$ character, since contributions from other orbitals near the Fermi
level are very small). We observe two B band systems: two bands are
derived from B $p_{z} $ states and four from B $p_{x,y}$. All
these bands are highly dispersive (light), the former being quite isotropic
and the latter more two-dimensional. Both $p_{z}$ bands cross the Fermi level
(in different parts of the Brillouin zone), but only 
 two bonding p$_{x,y}$ bands do so, and only near the $\Gamma$ point.
They form two small cylindrical Fermi surfaces
around the $\Gamma$-$A$ line (Fig.\ \ref{Fermi}).
However, due to their 2D character, they contribute more than 30\%
to the total $N(0)$.

In contrast, the $p_{z}$ bands have 3D character, 
since the smaller intraplane distance compensates for a smaller (pp$\pi $
$vs.$ pp$\sigma )$ hopping. 
In the nearest neighbor tight binding (TB) model their dispersion is
$\varepsilon _{{\bf k}}=\varepsilon_{0}+2t_{pp\sigma }
\cos ck_{z}\pm t_{pp\pi }\sqrt{3+2\cos {\bf a}_{1}{\bf k+}
2\cos {\bf a_{2}k+}2\cos {\bf a}_{3}{\bf k}}$ where {\bf a}$_{1,2,3}$
are the smallest in-plane lattice vectors. 
The on-site parameter $\varepsilon _{0}$ can be
found from the eigenvalue at the $K$ point and is $\sim 1.5$ eV above the
Fermi energy. We estimated 
$t_{pp\sigma }$ and $t_{pp\pi }$ from the
LMTO calculations as $\sim 2.5$ eV and $\sim $1.5 eV, respectively. This
model gives a very good description of the band structure near and below
the Fermi, although
the antibonding band acquires some
additional dispersion by hybridizing with the Mg $p$ band. The role
of Mg in forming this band structure can be elucidated by removing
the Mg atoms from the lattice entirely and repeating the calculations
in this hypothetical structure. The in-plane dispersion of the 
both sets of bands at and below the Fermi level changes very little
(for the $pp\pi$ bands there is hardly any change at all, while
the $pp\sigma$ in-plane dispersion changes by $\sim$ 10\%).
The $k_{z}$ dispersion of the $p_z$ bands is increased
in MgB$_2$ as compared with the hypothetical empty B$_2$ lattice
by about 30\%, and these bands shift down with respect to the $p_{x,y}$ bands
by approximately 1 eV. This shift, as well as the additional dispersion,
comes mainly from the hybridization with the empty
Mg $s$ band, which is correspondingly pushed
further up, increasing the effective ionicity. Substantial
$k_{z}$ dispersion of the  $p_z$ bands produces
the Fermi surface which is is approximately mirror-reflected with respect to
a plane between the $k_{z}=0$ and $k_{z}=\pi /c$ planes, one
pocket (electron-like) coming from the antibonding and the other (hole-like)
from the bonding $p_{z}$ band. The
two surfaces touch at one point on the K-H line and form a honeycomb
tubular network, replicating in
reciprocal space the boron lattice in real space.

The resulting bands are fairly 3D: the average Fermi velocities are 
$v_{x,y}=4.90\cdot 10^{7}$ cm/s and $v_{z}=4.76\cdot 10^{7}$ cm/s. The
plasma frequencies are $\omega _{px,y}=7.1$ eV and $\omega _{pz}=6.9$ eV.
Correspondingly, we predict fairly isotropic electrical resistivity, with
the linear slope $d\rho /dT\approx 0.08\lambda _{tr}$ $\mu \Omega 
$ cm$/K$. The total DOS at the Fermi level is $N(0)=0.36$
states/spin$\cdot $f.u.

The above described band structure is typical for an $sp-$metal. 
 What is {\it not }typical is
that this particular $sp-$metal is held together by covalent
bonding with a substantial ionic component, which
inevitably leads to a strong electron-phonon interaction. In the
following we present a semiquantitative estimate of the
corresponding coupling constant, and argue that the fortunate
combination of strong bonding, reasonable $N(0)$, and high phonon
frequency is responsible for the high transition temperature in this
compound.

In its most rigorous formulation the McMillan-Hopfield formula reads\cite
{problema}: 
\begin{equation}
\lambda =n^{-1}\sum_{ij}\left\langle NI_{i\alpha }I_{j\beta }\right\rangle
\left( \Phi ^{-1}\right) _{i\alpha ,j\beta },  \label{maks}
\end{equation}
where the indices $i,j$ run over all $n$ ions in the crystal; 
$\alpha , \beta $ are Cartesian indices; 
$\left\langle NI_{i\alpha } I_{j\beta }\right\rangle $
is the electron-ion matrix element averaged over the Fermi surface;
and $\Phi _{i\alpha ,j\beta }$ is the standard force matrix. While
this expression is exact, a number of simplifications are needed to make it
more practical. First, because of the large size of the Fermi surface, 
one can neglect the nondiagonal terms and reduce Eq.\ \ref{maks} to one unit
cell\cite{problema}: 
\begin{equation}
\lambda \approx \widetilde{\sum_{i}}\left\langle NI^{2}\right\rangle
_{i}\Phi _{ii}^{-1}=\widetilde{\sum_{i}}\eta _{i}\left( \Phi _{ii}\right)
^{-1}.
\end{equation}
The quantity 
$\Phi _{ii}=\partial ^{2}E_{tot}/\partial R_{i}^{2}$  is a local quantity, 
which can be calculated from the total energy differences of frozen
zone-center phonons. The Hopfield factor can be calculated in the Rigid
Muffin Tin (or Rigid Atomic Sphere) Approximation\cite{GG} (RMTA), assuming
that the change of the crystal potential due to an ion's displacement can be
described by shifting the  electronic charge distribution rigidly inside the
corresponding atomic sphere. Although expressions for $\eta $ have been
derived for arbitrary site symmetry\cite{mazsav}, these are complicated
and we will use here simplified formulas\cite{GG} formally
correct for cubic site symmetry. With this simplification, $\eta $ can be
readily calculated from the LMTO potential parameters and partial DOS's.
We show the results in Table \ref{t1}.

In order to get an estimate of the phonon spectrum of the system, 
we calculated all zone-center modes, using the full-potential LAPW method. 
There are four distinct modes  \cite{iso}:
one silent mode, $B_{1g}$ (two borons displaced along $z$
in opposite directions), one doubly degenerate Raman mode, $E_{2g}$
(in-plane displacements of borons), and two infrared-active modes, which do
not involve changes of in-plane bonds: $A_{2u}$ (B and Mg planes
moving against each other), and a doubly degenerate $E_{1u}$ mode (B and Mg
planes sliding along $x,y).$ Their frequencies are, respectively, 690 cm$
^{-1},$ 470 cm$^{-1}$, 390 cm$^{-1}$, and 320 cm$^{-1},$ and their force
constants are 390, 180, 70, and 44 mRy/$a_{B}^{2}$/atom. 
All calculations were carried out at the experimental lattice constant
and $c/a$ ratio \cite{exp-latt}.

Given the physics of the electronic structure described above, 
it seems likely that the lowest mode couples little with the electrons 
and that its softness is derived from its in-plane acoustic character. 
Thus, we excluded it from the calculations of $\lambda $
below. Correspondingly, the average inverse force matrix for boron is 
$\left( \Phi _{B}\right) ^{-1}=M_{B}^{-1}\left\langle \omega
^{-2}\right\rangle =7.1 a_{B}^{2}$/Ry, corresponding to $\left\langle
\omega ^{-2}\right\rangle ^{-1/2}\approx 400$ cm$^{-1}$. Together with the
above value for $\eta $ this gives $\lambda \approx 0.7$ and the
logarithmically averaged frequency of the same three modes is 
$(690\cdot 470^2 \cdot 390)^{0.25}=\left\langle
\omega _{\log }\right\rangle \approx 500$ cm$^{-1}\approx 700$ K. We can now
estimate the critical temperature according to the McMillan formula, $T_{c}=
\frac{\left\langle \omega _{\log }\right\rangle }{1.2}\exp [-1.02(1+\lambda
)/(\lambda -\mu ^{\ast }-\mu ^{\ast }\lambda )].$ Using for the Coulomb
pseudopotential $\mu ^{\ast }$ the commonly accepted value 0.1, 
we obtain $T_{c}\approx $22 K.

It is instructive to compare the calculation above with a typical $sp$
superconductor, Al (Table \ref{t2}). RMTA notoriously underestimates the
electron-ion scattering in $sp$ metals due to the weaker screening
than in $d$-metals. Also in Al, $\lambda$ is underestimated by a factor of 2. 
The screening properties of MgB$_{2}$ are similar to those of
Al (they have similar Thomas-Fermi screening lengths), and the electronic
properties are similar as well. It is therefore
tempting  to scale the calculated $\lambda$ for MgB$_{2}$  by the same factor. 
The scaled results are also shown in Tab.\ \ref{t2} and the corresponding
$T_{c}$ is approximately 70 K. 
Although this table looks impressively quantitative, these
numbers should be regarded only as rough estimates.
The  RMTA in MgB$_{2}$ is
clearly a worse approximation than in Al. The boron site symmetry is far from
cubic and, moreover, large differences in the atomic sphere radii lead to
artificial potential jumps at the sphere boundary, a problem
for which there is no remedy in RMTA. 
In other words, with regard to EPC, our calculations
should be considered as a qualitative indication of a strong electron-phonon
interaction. We can nevertheless be confident of our main
qualitative conclusions.

Our main conclusion is that MgB$_{2}$, being essentially a metallic boron held
together by covalent B-B and ionic B-Mg bonding, is electronically
a typical $sp$ metal with a typical DOS. Strong bonding induces strong
electron-ion scattering and hence strong electron-phonon coupling. An
additional benefit is the high frequency of the boron vibrations (while the
force constants remain reasonably soft). Superconductivity is mainly due to
boron. The light mass and correspondingly large zero-point vibrations 
($>0.1 a_{B}$) suggest a possibility of anharmonic and/or nonlinear EPC, and
possible deviation of the isotope effect from 100\%. Isovalent doping may be
beneficial if it increases the density of states $N(0)$. Lattice expansion
due to Ca doping should lead to an overall increase of the density of
states, and may provide the additional benefit of reduced $p_{z}$-$s$-$p_{z}$
hopping. Another interesting dopant is Na, which should not only expand the
lattice, but also decrease the Fermi level, exposing more of the $p_{x,y}$
bands, which may provide an additional contribution to $\lambda$.

Finally, let us outline the directions for further theoretical
investigation. First, EPC calculations beyond the RMTA ({\it e.g.,} in the
linear response formalism) are highly desirable and computationally
feasible. Second, calculations with full structure optimization for
(hypothetical) CaB$_{2}$ and BeB$_{2},$ and virtual crystal calculations for Na
doping should elucidate the effect of isovalent and hole doping, giving some
hints toward further optimizing $T_{c}$. Work along these lines is
currently in progress.

We are thankful to J.E. Pask, C.S. Hellberg and D.J. Singh
for critical reading of the manuscript. This research was supported in
part by ONR and by DOE under Contract No. W-7405-82.

\begin{table}
\caption{Partial LMTO DOS, $N(0)$, in eV$^{-1}$/spin
and partial Hopfield factors, $\eta,$  in mRy/$a_B^2$,
for Mg and B (per atom).}
\begin{tabular}{ccccc|ccc}
& \multicolumn{4}{c}{$N(0)$} &\multicolumn{3}{c}{$\eta$} \\ 
$l$ & $s$   & $p$   & $d$   & $f$   & $sp$ & $pd$ & $df$ \\ 
\tableline
Mg  & 0.018 & 0.043 & 0.078 & 0.019 & $<$1 & 2    & 3 \\ 
B   & 0.003 & 0.199 & 0.010 & -     & $<$1 & 135 & - \\
\end{tabular}
\label{t1}
\end{table}
\multe
\begin{table}
\caption{Comparison of the electron-phonon coupling
in Al and in MgB$_2$. The entries labeled by a question mark
are obtained by scaling the RMTA $\lambda$ by the ratio
of $\lambda_{exp}/\lambda_{rmt}$ in Al. $q_{TF}$ is the Thomas-Fermi
screening parameter in $a_{B}^{-1}$. 
$T_c$ was calculated by the McMillan formula with $\mu^*=0.1$.
LAPW $N(0)$ is given in eV$^{-1}$/spin, $\eta$ in mRy/$a_B^2$
 and $\left\langle M\omega^{2}\right\rangle $ in Ry/$a_{B}^{2}$.  }
\begin{tabular}{ccccccccccc}
& $N(0)$ & $\eta _{rmt}$ & $\left\langle 
\omega _{\log }\right\rangle $ & $\left\langle M\omega
^{2}\right\rangle $ & $\lambda _{r.m.t.}$  & $T_{c}(\lambda _{rmt})$ 
& $\lambda _{\exp }$ & $T_{c}(\lambda _{\exp })$ & $T_{c,\exp }$ & $q_{TF}$  \\
\tableline
Al & 0.15 & 27 & 250 & 0.135  & 0.2 & 0 & 0.4 & 1.3 & 1.3 & 0.73
\\
MgB$_{2}$ & 0.18 (per B) & 135 & 500 & 0.141 & 0.7 & 22 & 1.4?
& 70? & 39 & 0.70
\end{tabular}
\label{t2}
\end{table}
\eject
\multb

%\multe \multb
\vskip -1cm
\multe
\end{document}